# Direct Observation of Giant Saturation Magnetization in $Fe_{16}N_2$

Nian Ji[1,2]*, Valeria Lauter[3,]*, Cheng-Jun Sun[4], Lawrence F. Allard[5], Hailemariam Ambaye[3], Steve M Heald[4], Edgar Lara-Curzio[5], Xiaoqi Liu[1], Yunhao Xu[1], Xuan Li[1] and Jian-Ping Wang[1,2 #]

1. *The Center for Micromagnetics and Information Technologies (MINT) and Department of Electrical and Computer Engineering, University of Minnesota, Minneapolis, Minnesota 55455, USA*

2. *School of Physics and Astronomy, University of Minnesota, Minneapolis, Minnesota 55455, USA*

3. *Neutron Scattering Science Division, Oak Ridge National Laboratory, Oak Ridge, TN 37831 USA*

4. *Advanced Photon Source, Argonne National Laboratory, Argonne, IL 60439, USA*

5. *Materials Science & Technology Division, Oak Ridge National Laboratory, Oak Ridge, TN 37831, USA*

*\* Authors contributed equally*

*# To whom correspondence should be addressed. email: jpwang@umn.edu*

## Abstract

Magnetic materials with giant saturation magnetization ($M_s$) have been a holy grail for magnetic researchers and condensed matter physicists for decades because of its great scientific and technological impacts. As described by the famous Slater-Pauling (SP) curve[1,2], the material with highest $M_s$ is the $Fe_{65}Co_{35}$ alloy ($M_s \sim 1900$ emu/cm$^3$)[3]. This was challenged in 1972 by a report on the compound $Fe_{16}N_2$ with $M_s$ about 18% higher than that of $Fe_{65}Co_{35}$[4]. Following this claim, there have been enormous efforts to reproduce this result[5, 6, 7, 8, 9, 10, 11, 12, 13] and to understand the magnetism[14, 15, 16, 17, 18] of this compound. However, the reported $M_s$ by different groups cover a broad range[19], mainly due to the unavailability of directly assessing $M_s$ in $Fe_{16}N_2$. In this article, we report a direct observation of the giant saturation magnetization up to 2500 emu/cm$^3$ using polarized neutron reflectometry (PNR) in epitaxial constrained $Fe_{16}N_2$ thin films prepared using a low-energy and surface-plasma-free sputtering process. The observed giant $M_s$ is corroborated by a previously proposed "Cluster + Atom" model[20], the characteristic feature of which, namely, the "directional charge transfer" is evidenced by polarization-dependent x-ray absorption near edge spectroscopy (XANES).

**Introduction**

While the material α"-$Fe_{16}N_2$ and its interesting magnetic behavior were discovered decades ago, there is still no unified answer to the question on whether this phase has a giant saturation magnetization ($M_s$). An important aspect that emerges upon examination of the magnetization of this system is that conventional magnetometer-based (VSM or SQUID) methods can only measure the total magnetic moment of the samples. Given the metastable nature of α"-$Fe_{16}N_2$, it has been challenging to develop bulk samples with pure phase (N fully ordered)[21]. For thin film processing, to avoid the formation of thermodynamic equilibrium phases (e. g. γ'-$Fe_4N$), "epitaxial constraint" is a method of choice. However, these films usually require complicated choices and delicate treatment of underlayers and substrates. The subsequent evaluation of the $M_s$ values of these thin films involves challenging estimation of thin film volume and subtle assessment of magnetic contributions from underlayers, substrates or possible impurity phases, resulting in unpredictable errors. These issues have been the primary cause of the controversy of this topic since the original claim of giant saturation magnetization 40-years agoError: Reference source not found.

The polarized neutron reflectometry (PNR) measurement method has become increasingly popular as an experimental technique to probe the depth-dependent magnetic properties of thin films over the last decades, due to the availability of more intense sources[22]. Given the sensitive interaction between the neutron spin and magnetic inducting field (B) within the ferromagnetic materials, it allows a *direct* determination of the depth-dependent magnetic structure[23]. Inspired by these considerations, we used PNR to study epitaxial $Fe_{16}N_2$ thin films, and observed a giant $M_s$ up to 2500 emu/cm$^3$, providing direct evidence of the existence of giant $M_s$ in this material system.

## Results

### Structural analysis of epitaxial Fe-N films

We start our discussion by providing a detailed structure and chemical analysis. Epitaxial $Fe_{16}N_2$ thin films are grown on GaAs (001) single crystal substrate by using a facing target sputtering (FST) process[24, 25]. A typical out-of plane $\theta/2\theta$ scan is shown in Fig. 1a, which clearly indicates the formation of the chemically ordered $Fe_{16}N_2$ phase with (00l) orientation as evidenced by the signature (002) reflection coming from the N site ordering. Remarkably, as explored by the grazing incident x-ray diffraction (GIXRD) (Fig. 1b and c), the in-plane lattice constant substantially deviates from its bulk value. Aligning the scattering vector along the GaAs (220) and GaAs (400), the dominant film peaks at 40.8° and 58.6° can be indexed to $Fe_{16}N_2$ (220) and (400) respectively with a calculated in-plane lattice constant of 6.26 Å, which is ~10% larger than its bulk value. The corresponding $\Phi$ scan (Fig. 1d) performed on these diffraction peaks show four regularly spaced peaks at 90° spacing, suggesting an expected 4-fold cubic symmetry given the tetragonal lattice of $Fe_{16}N_2$. Since a thin Fe is deposited on the surface-untreated GaAs at an elevated temperature, the enhancement of the in-plane lattice constant can be attributed to the formation of an Fe/GaAs interfacial phase, likely $FeGa_3$, (crystallizes in tetragonal space group: *P42/mnm*, a=6.26 Å, c=6.55 Å), which promotes the epitaxy with larger lattice constant and forces the in-plane lattice to stretch measurably. In common rigid metals, straining of crystal lattice by coherent growth is limited to ultrathin films with thicknesses of only up to several atomic layers due to the requirement of substantial elastic energy[26]. In the case of ferromagnetic martensite (Fe-N martensite here) because the energy scale is relatively flat over the entire Bain path[27], it is possible to stabilize the intermediate lattice geometry over a wide thickness range. To further verify the structural information, we have measured x-ray reflectivity in one annealed sample in fig. 1e with the depth-profile showing in the inset. It is clear that the bulk part of the samples shows a scattering length density of $4.6 \times 10^{-5}$ Å$^{-2}$, which is much less than that of nominal $Fe_{16}N_2$ ($5.9 \times 10^{-5}$ Å$^{-2}$), suggesting a significantly reduced mass density due to in-plane lattice expantion, which is consistent with the in-plane x-ray diffraction analysis.

Fig. 2a shows a high-angle annular dark-field scanning transmission electron microscopy (HAADF-STEM) image from a cross-sectional multilayer sample. The layers show brighter contrast for higher average atomic number, so Fe, for example, is the brightest layer in this image. Both microbeam and selected-area diffraction patterns (SADP) were recorded as appropriate in different regions (Fig. 2b~e); they correspond in general position to the labeled areas in the HAADF image. These diffraction results reveal the overall crystalline quality of the film and the epitaxy of the lattice structure of the Fe-N layer, which is highly coherent with the Fe buffer layer. With the sample oriented to the [110] zone axis of the GaAs substrate, the α-Fe layer is in a precise [110] orientation. Using the exact d-spacings for the Fe reflections to calibrate a camera factor for further indexing of the rest of the pattern, the d-spacings for the FeN layer are shown to closely match a [1-10] zone axis of Fe lattice that has the same tetragonal space group (*I4/mmm*) as the stoichiometric $Fe_{16}N_2$. A microbeam diffraction pattern with the beam spread to cover a 75 nm diameter area, was recorded, covering primarily the Fe and FeN layers, with a small contribution from the Ir capping layer and the C layer deposited by the focused ion beam (FIB) process. To bring out additional diffraction information, the contrast of this pattern was expanded as shown in Fig. 2f (the as-recorded image is shown in Fig. S1). Faint diffuse rings from the amorphous C layer are seen, and a few sharp, irregularly spaced reflections from the crystals in the as-deposited Ir capping layer are also present. The predominant feature of the pattern is, however, the periodic array of diffuse reflections that match the ordering reflections of the $Fe_{16}N_2$ compound. These reflections are circled in Fig. 2f. The measured spacings and indexing of the diffuse pattern is consistent with the computed [1-10] $Fe_{16}N_2$ pattern shown in Fig. 2g. Bright-field STEM imaging and the corresponding energy-dispersive x-ray mapping for each constituent element (Fig. 3) reveal that N and Fe atoms were distributed homogeneously inside the thin-film region without any segregation at the surface or interface, which is consistent with an Auger electron spectroscopy measurement (Ref. Error: Reference source not found).

**Polarized neutron reflectometry measurement**

We utilized the technique of polarized neutron reflectometry (PNR) to directly explore the depth-dependent magnetization profile of the present $Fe_{16}N_2$ thin films[28]. As shown in Fig. 4a, the spin-polarized (μ+ for spin-up and μ- for spin-down) neutron beam impinges onto the sample surface at a grazing incident angle $\alpha_i$ and specularly reflects at an angle $\alpha_f$ ($\alpha_i = \alpha_f$). The reflectivity data for the two spin states are recorded as functions of momentum transfer $q_z = 4\pi \sin \alpha_i / \lambda$, with $\lambda$ being the neutron wavelength. In this case, the propagation of the neutron beam can be presented by optical formalism[29, 30], in which the interaction between the radiation and the medium is described by the Fermi pseudo potential[31].

$$V_{\pm} = \frac{2\pi\hbar^2}{m_n} N(b_n \pm b_m),$$

where $\pm$ denotes the spins of neutrons parallel and anti-parallel to the external field H, respectively; N represents number density of the material ($\sim$Å$^{-3}$), $m_n$ neutron mass, $b_n$ and $b_m$ denote the nuclei and magnetic scattering amplitude ($\sim$Å), respectively. From the fits to the two reflectivity curves (R$^+$ and R$^-$), the depth-dependent nuclear (N$b_n$) and magnetic (N$b_m$) scattering length density profiles (NSLD and MSLD), respectively, are determined. In particular, when measuring at magnetically saturated state, the MSLD (Nbm) obtained is linked to Ms[32] and allows its direct evaluation.

The PNR experiments were performed using the Magnetism Reflectometer on beamline 4A at Spallation Neutron Source, Oak Ridge National Laboratory[33]. This is a time-of-flight instrument with a wavelength band of 2–5 Å and a polarization efficiency of 98%. The PNR experiments were performed at room temperature in a saturation external magnetic field of H = 10 kOe applied parallel to the thin film plane. The non-spin-flip (NSF) intensities of spin up and spin down neutrons were recorded alternatively by a position sensitive detector.

We start our PNR discussion by first analyzing the experimental results from a reference sample of *pure* Fe film on GaAs substrate prepared by the same FST method as Fe-N samples. Fig. 5b shows the experimental reflectivities and the corresponding simulations. From the fit to the data we obtained both NSLD and MSLD being close to bulk Fe. This confirms the expected result that the thin Fe film possesses uniform magnetization, which reasonably agrees with that of bulk Fe ($\sim$1700 emu/cm$^3$).

For the Fe-N films, the PNR experiments were performed under the same conditions as the reference Fe film. The results obtained for three samples with different annealing time treatment are shown in the Fig. 5a-c along with corresponding fits to the data. Looking at the reflectivity curves of each sample, in addition to the difference of the periodicity of the fringes due to the thickness variation, it is noticed that as the annealing time progressively increases, the oscillation magnitude gradually evolves, which is especially robust for the $R^-$ curve. In particular, for the as-grown sample, the reflectivities show pronounced oscillations in contrast to that of the 20 hrs-annealed sample, where the reflectivities become very smooth and without well-defined oscillations. Because of this behavior, fits to the data reveal difference in the depth-dependent magnetic properties of these films.

The middle panel, Fig. 5d-f shows the NSLD depth-profile as functions of the distance from the substrate of the four samples, the configurations of which is consistent with the structural analysis by XRD (Fig. 1) and are rather uniform through the thickness except a "bump" at the bottom interface between film and substrate, corresponding to the seed Fe layer.

The striking results of the MSLD profiles are shown in the bottom panel of Fig. 5g-i. For all the annealed samples, the $M_s$ substantially exceed that of $Fe_{65}Co_{35}$ (~1900 emu/cm$^3$ outlined by dashed blue line). As we progressively increase the annealing time, the $M_s$ of the bulk part of the sample enhances significantly and uniformly, reaching up to $M_s$ ~ 2500 emu/cm$^3$. In our previous report (Ref.Error: Reference source not found), it is observed that the degree of N site ordering is correlated with the annealing time according to XRD analysis. Therefore, the obtained results prove that the enhancement of $M_s$ after a longer annealing time is due to the increasing amount of chemically ordered $Fe_{16}N_2$ phase, which is consistent with the previous magnetometry analysis.

Furthermore, we compared our neutron results with the bulk magnetization data obtained using VSM with the error estimation discussed explicitly in the methods section. As shown in Fig. 6, we have tested five samples of pure Fe film, as-prepared Fe-N film and three Fe-N films with different annealing time treatment. The extracted magnetizations from both measurements are in good agreement.

**Discussion**

Now we turn to the origin of this giant $M_s$. In our previous studies, we have proposed a possible scenario to realize a giant $M_s$ beyond the Slater-Pauling curve (Ref. Error: Reference source not found). An immediate predication of the "Cluster ($Fe_6N$) + Atom (Fe)" model is the presence of directional charge transfer from Fe to N site, in which case, an effective "double exchange" mechanism can be facilitated and consequently, a high-spin configuration and long-range ferromagnetic order can be stabilized to develop high moment (Fig. 7a). To verify this essential feature, we investigated the nature of Fe-N bond from individual Fe sites using polarization-dependent x-ray absorption near edge spectroscopy (XANES), which allows the evaluation of the local structure with element specificity[34, 35, 36, 37]. The glancing angle XANES measurements were performed with the x-ray polarization parallel (E//surface) and perpendicular (E⊥surface) to the $Fe_{16}N_2$ (002) surface plane (Fig. 7b). The analysis was done by comparing the simulated spectra with ($Fe_{16}N_2$) and without ($Fe_{16}N_0$) N occupying the octahedral center to extract the role of N using the *ab initio* FEFF8.4 code[38, 39]. The simulated polarization-dependent XANES based on two structural models of $Fe_{16}N_2$ and $Fe_{16}N_0$ are shown in Fig. 7c and d, respectively. The simulated XANES spectra of $Fe_{16}N_2$ reproduce all essential features as observed experimentally, as marked using the vertical blue dashed lines, which is in contrast to that of $Fe_{16}N_0$. Therefore, we used the simulated XANES spectra from individual Fe sites to investigate the role of N in the $Fe_{16}N_2$. As shown in Fig. 6e and f for the E//(002) surface plane, the only substantial spectra modification upon removing N is seen on Fe8h (Fig. 6g), which is attributed to its large influence on the in-plane Fe-N bonds. In comparison, the XANES spectrum from the Fe4e site shows a slight change (Fig. 7f) for the E//(002) surface plane direction. Likewise, a similar conclusion can be drawn for the case of Fe4e and the E⊥(002) surface plane, which results from the out-of-plane Fe-N bonds. The Fe4d site, which is the most distant Fe atom from the N site, undergoes the least spectra modification with N occupation status change. Most importantly, it is found that the absorption edges of Fe4e for E⊥(002) surface plane and Fe8h for E//(002) surface plane with N ($Fe_{16}N_2$) consistently shift to higher absorption energies as compared to absorption edges of $Fe_{16}N_0$. These polarizations correspond to the directions of Fe-N bonds. It is well established that the absorption edge is consistently shifted to higher energies with higher Fe valance states (Fig. S2), indicating that the energy shift in the Fe absorption edge is associated with the charge transfer from Fe to N.

To summarize, we have directly measured and confirmed the existence of the giant saturation magnetization in $Fe_{16}N_2$ films utilizing PNR technique. The charge transfer that predicted by the "cluster + atom" model was supported by the XANES observation, which provides a new insight on the origin of the giant $M_s$ in $Fe_{16}N_2$.

**Methods**

**Synthesis of $Fe_{16}N_2$ films**

The iron nitride samples for the PNR experiments were prepared by first depositing a thin Fe layer with nominal thickness of 2 nm on GaAs (001) single crystal substrate at an ambient temperature of ~300 °C. The Fe-N layer was subsequently grown by sputtering thoroughly mixed Ar and $N_2$ gases at room temperature. The x-ray diffraction (XRD) and transmission electron microscopy studies reveal the formation of body-centered tetragonal Fe-N martensite epitaxially adopted on GaAs(001). Fine-tuning of the $N_2$ doping rate results in the formation of stoichemtry Fe/N=8/1 as feedback from Auger electron depth profile. After an *in-situ* post-annealing treatment at substrate temperature of 120 °C on the as-grown samples, chemically-ordered $Fe_{16}N_2$ phase can be favored with its degree of N site ordering controlled by the amount of annealing time, which is varying from 5hrs to 20 hrs.

**Electron microscopy sample preparation and imaging**

Samples for analysis via methods of aberration-corrected high-resolution electron microscopy and electron diffraction were prepared from small pieces of the sample using focused ion-beam (FIB) milling techniques. The FIB samples were examined in a Hitachi HF-3300 cold-field-emission TEM operated at 300 kV, for bright-field (BF) imaging and for electron diffraction pattern acquisition by selected-area-diffraction and microbeam diffraction methods. In the latter case, a 10 μm condenser aperture was used and the incident beam set at a diameter of 30 to 75 nm to restrict the area from which diffraction patterns were generated; the larger-diameter beam conditions yielded convergent-beam patterns with reflections sharp enough for precision measurements of d-spacings. Images and diffraction patterns were acquired using a Gatan Ultrascan Model 1000 multi-scan CCD camera with a 2k x 2k imaging chip.

**$M_s$ measurement using Magnetometry**

A Vibrating Sample Magnetometer (VSM) with a sensitivity in the range of $\sim 10^{-7}$emu with applied field up to 10kOe at room temperature was used to measure the total magnetic moment. To measure the volume of the samples, thickness was measured by X-ray reflectivity (XRR) and selectively cross-checked by cross-section TEM image to ensure the accuracy. The error bar is typically 2% determined from the number of visible fringes in the XRR curve. To evaluate the film area, two methods were applied to reduce the error: 1) Samples subject to VSM test were cut into rectangular shape, the total area were calculated by measuring the sidelength using micrometer. 2) The total mass of the same sample was measured and compared with the weight of a reference GaAs substrate with known area. The difference of the obtained value of the sample area between these two methods were treated as the error bar which is <2%. The absolute value of magnetization typically falls in the range of $10^{-3} \sim 10^{-4}$emu with testing error $\sim 3\%$. Standard Ni sample with known magnetic moment (44memu) was tested prior to each measurement to ensure the accuracy. It should also be mentioned that the VSM results presented here are not from a single point measurement. In fact, two or more samples cut from the same piece were checked. The scatter in the magnetization values was within 5%.

**Acknowledgement**

The work done at UMN and APS at ANL are supported by Office of Basic Energy Sciences under No. DE-AC02-98CH10886 and No. DE-AC02-06CH11357. Parts of this work were carried out through the support from NSF through the MRSEC program (DMR-0819885). The microscopy work done at HTML and PNR work done at SNS at ORNL are sponsored by U.S. Department of Energy, Office of Energy Efficiency and Renewable Energy, Vehicle Technologies Program. The authors would like to thank the useful conversations with Prof. P. Crowell. The authors also thank Prof. J. Judy, Prof R. H. Victora and Prof. R. James for proof-reading the manuscript.

**Author Contribution**

J.-P. W initiated and coordinated the overall work. N. J. and J.-P. W synthesized the samples and carried out the XRD, XPS and magnetic measurement. V. L., H. A. and N. J. performed the PNR measurement, C. J. S. and S. H. performed the XANES measurement, L. F. A and E. L. -C. performed the HR-TEM work. J.-P. W proposed the model and N. J. X. Q. L. contributed to the $Fe_6N$ cluster model analysis. Y.H. X and X. L. contribute to the sample preparation at the beginning of the project. N. J., V.L. and J.-P. W drafted the paper. All authors contributed to the data analysis and paper writing.

**Figure Captions**

**FIG. 1 x-ray diffraction characterization a** Typical high angle x-ray diffraction data on a $Fe_{16}N_2$ thin film sample. **b** and **c** In-plane X-ray diffraction with scattering vector aligned with GaAs (220) and GaAs (400) respectively, corresponding to a d-spacing of 6.26Å (Peaks labeled with * is due to the surface oxides)  **d** the in-plane Φ scan on diffraction peaks of (220) and (400). The inset show the zoom-in look of (220) peak as marked by the arrow. **e** X-ray reflectivity measured on one $Fe_{16}N_2$ sample with the fitted model shown in the inset.

**FIG. 2 HAADF-STEM image and diffraction patterns of epitaxial $Fe_{16}N_2$ cross-sectional sample.** A HAADF image of a layered Fe-N sample. **b-e** diffraction patterns from the general positions located by the outlined areas in **a**. **f** contrast expanded microbeam diffraction pattern showing [110] zone-axis. The periodic array of reflections consistent with the ordering reflections shown in **g**, the indexed computed pattern for the $Fe_{16}N_2$ structure.

**FIG. 3 Chemical analysis of Fe-N layers.** Bright field image of a cross-section sample and the corresponding energy-dispersive x-ray spectroscopy mappings for Fe, N, Ga, As and Ir. The presence of a uniform distribution of N in the Fe-N layer is clearly evident.

**FIG. 4 Polarized neutron reflectometry (PNR). a** A sketch of the scattering geometry of the polarized neutron reflectometry experiment. The spin-up (μ+) and spin-down (μ-) neutrons are defined as spins parallel and anti-parallel to the external magnetic field H, so that reflectivity R+ and R- probe the magnetization (M) direction of the sample. **b** The PNR data (R+ and R- ) with fitted curves of a reference Fe film grown on GaAs substrate **c** The nuclear and magnetic depth-profiles of the  scattering length density (NSLD: $Nb_n$

and MSLD: $Nb_m$, respectively) shown as functions of the distance from the substrate obtained from the fitting in (b). The blue dashed line represents the reference $M_s$ value of $Fe_{65}Co_{35}$ for comparison.

**FIG. 5 PNR analysis on Fe-N samples with giant $M_s$ a~c** Measured PNR reflectivities obtained from Fe-N samples after different annealing time treatment as outlined in the figure, along with fitting curves to the data. . **d~f** Nuclear and magnetic **g~i** scattering length density depth-profiles as functions of the distance from substrate for each film after different annealing time. The MSLD is presented on the left hand side and the magnetization in cgs unit ($emu/cm^3$) is shown on the right hand side of the figures. The blue dashed line represents the reference $M_s$ value of $Fe_{65}Co_{35}$ for comparison.

**FIG. 6 Magnetic characterization comparison**. Comparison between VSM and PNR results on average saturation magnetization value obtained for five different samples: pure Fe film, as prepared Fe-N films, three Fe-N after different annealing time. The blue line plots the ideal case when the two measurements (PNR and VSM) yield the same results.

**FIG. 7 Experimental proof of "cluster + atom" model to rationalize high $M_s$. a** An illustration of the "atom+cluster" model to realize the giant moment in $Fe_{16}N_2$, within the $Fe_6N$ octahedron, the charge transfer facilitates the high spin configuration and double exchange FM coupling. Exterior the $Fe_6N$ cluster, the intermediate spin state is energetically favorable. XANES analysis of $Fe_{16}N_2$ experimental **b** $Fe_{16}N_2$ calculated **c** and $Fe_{16}N_0$ calculated **d** spectra with polarization direction of X-ray perpendicular (red) and parallel (black) to the sample surface. **e** The crystal structure of $Fe_{16}N_2$ and each individual site. The calculated XANES spectra on of $Fe_{16}N_2$ and $Fe_{16}N_0$ on both Fe4e and Fe8h sites are shown in **f** and **g**, respectively. The horizontal dash blue line refers to the normalized absorption value as 0.5. The absolute K-edge of XANES simulation of Fe metal is shifted 10.5 eV higher compared to that of Fe metal XANES measurement.

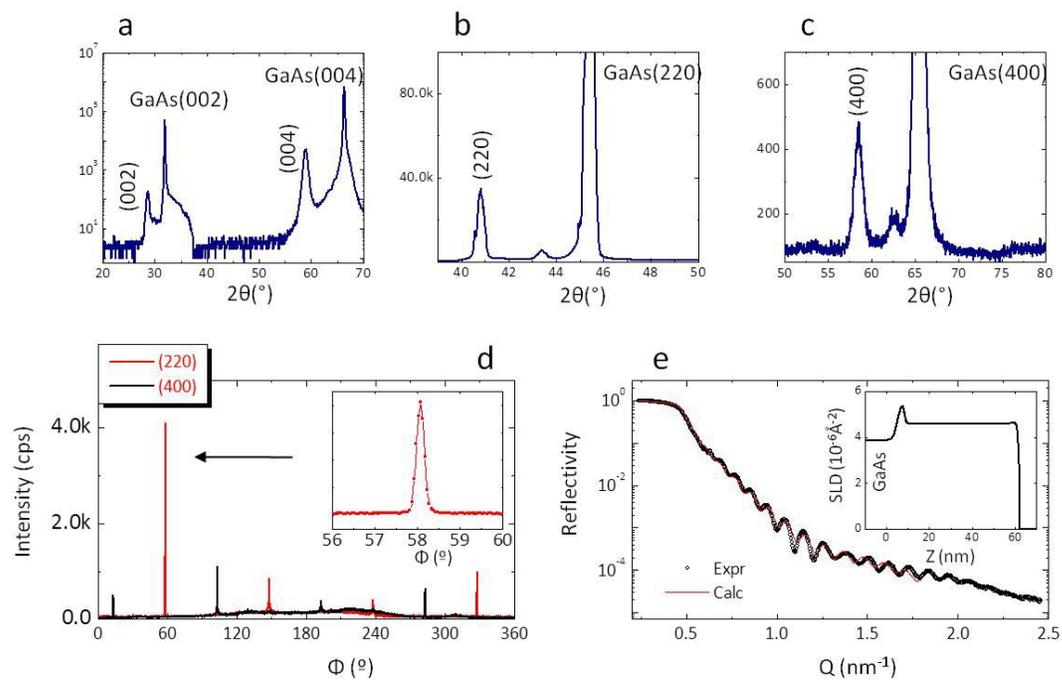

FIG.1

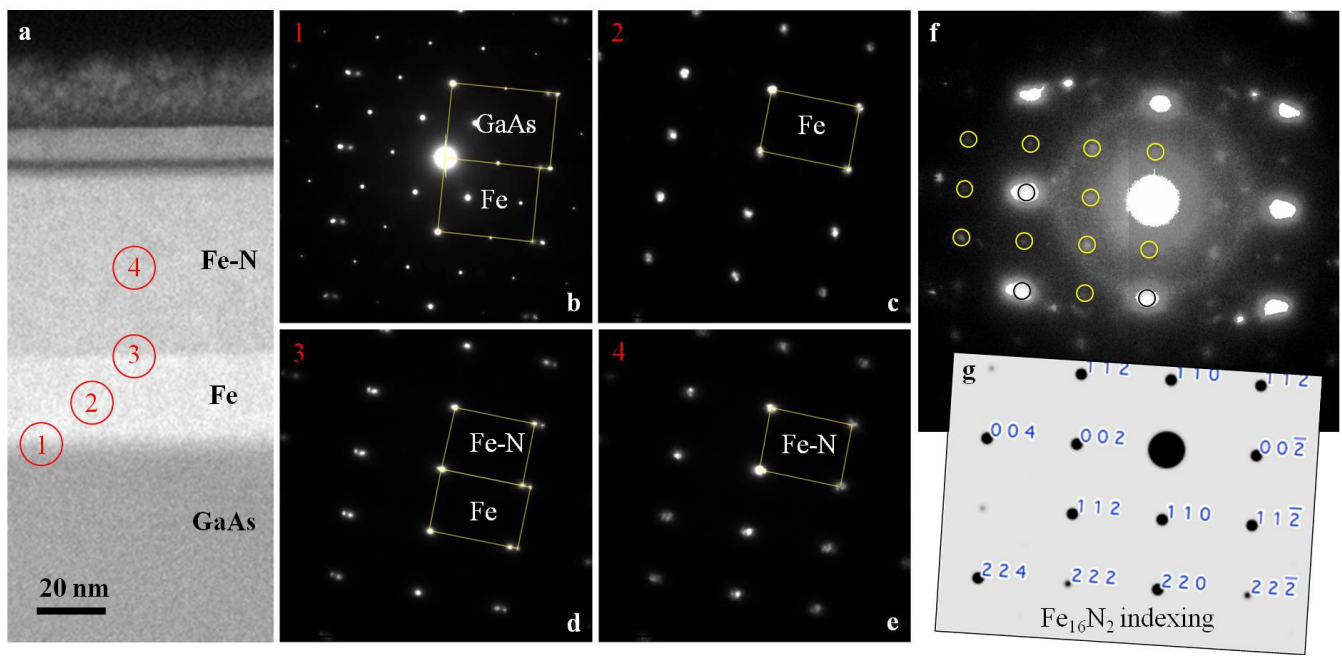

FIG 2

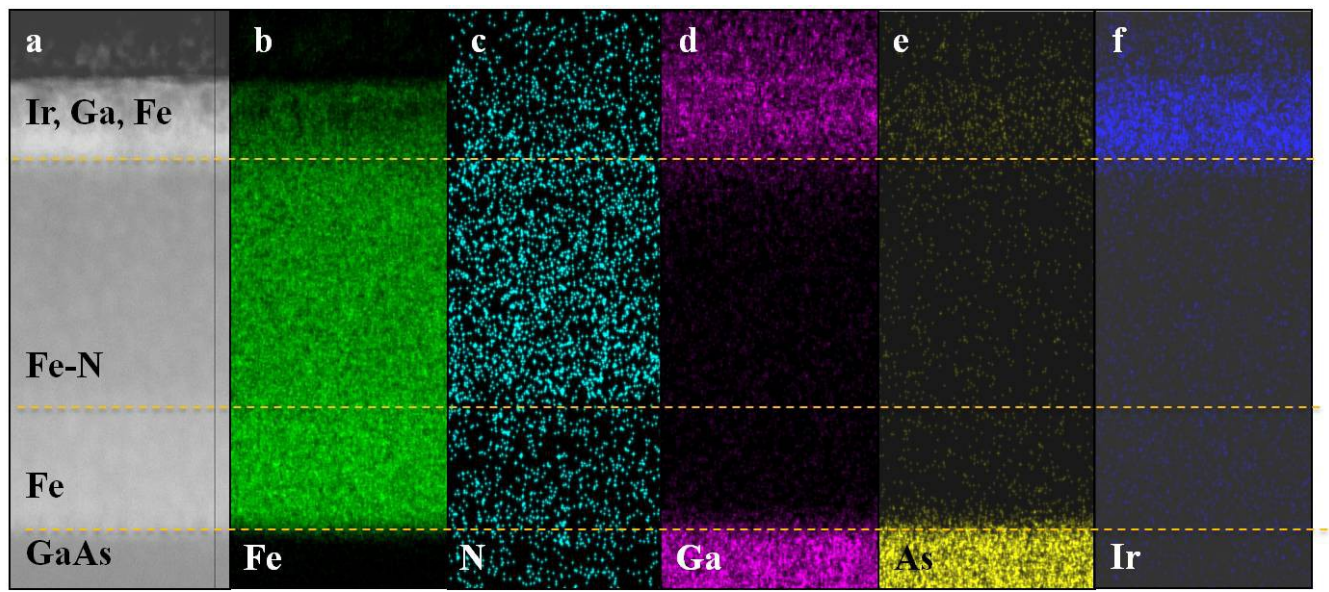

FIG. 3

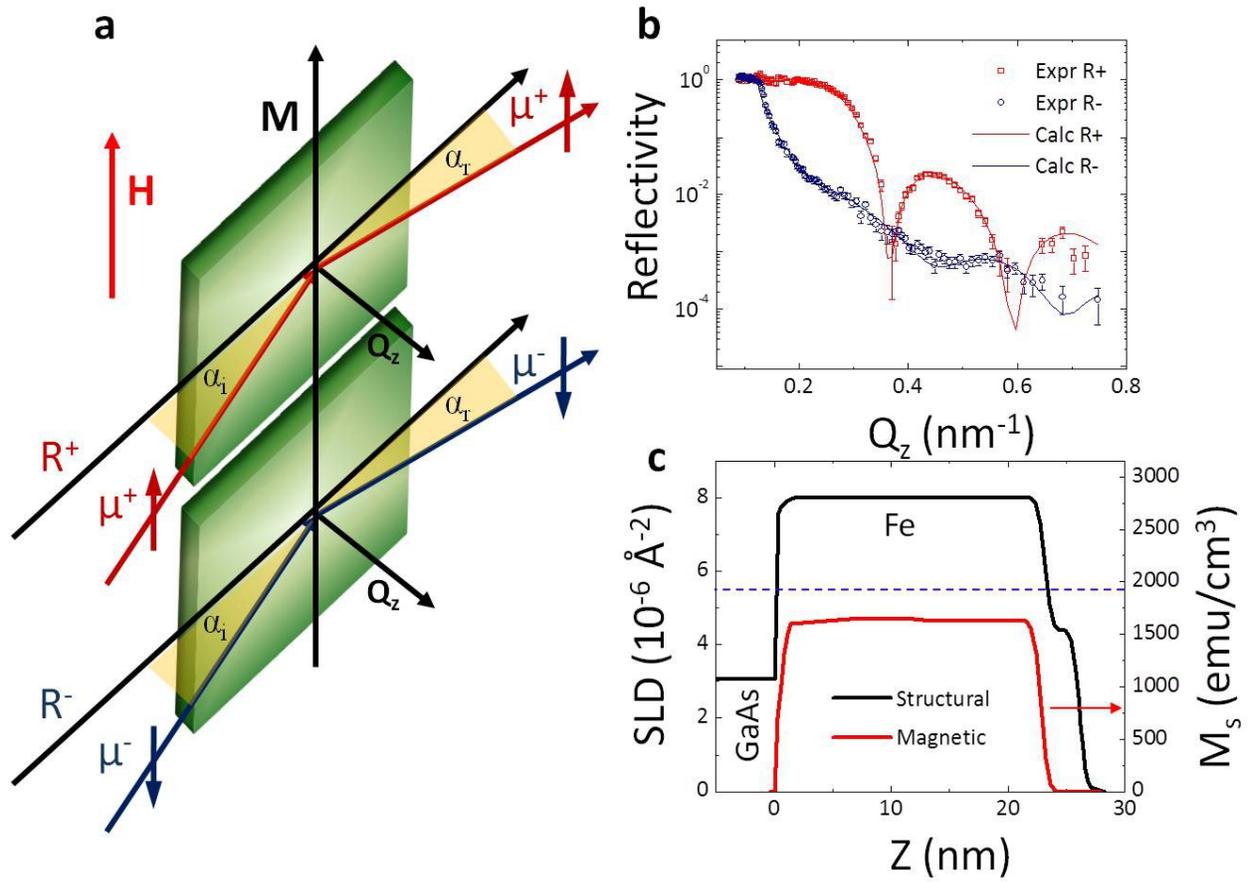

FIG. 4

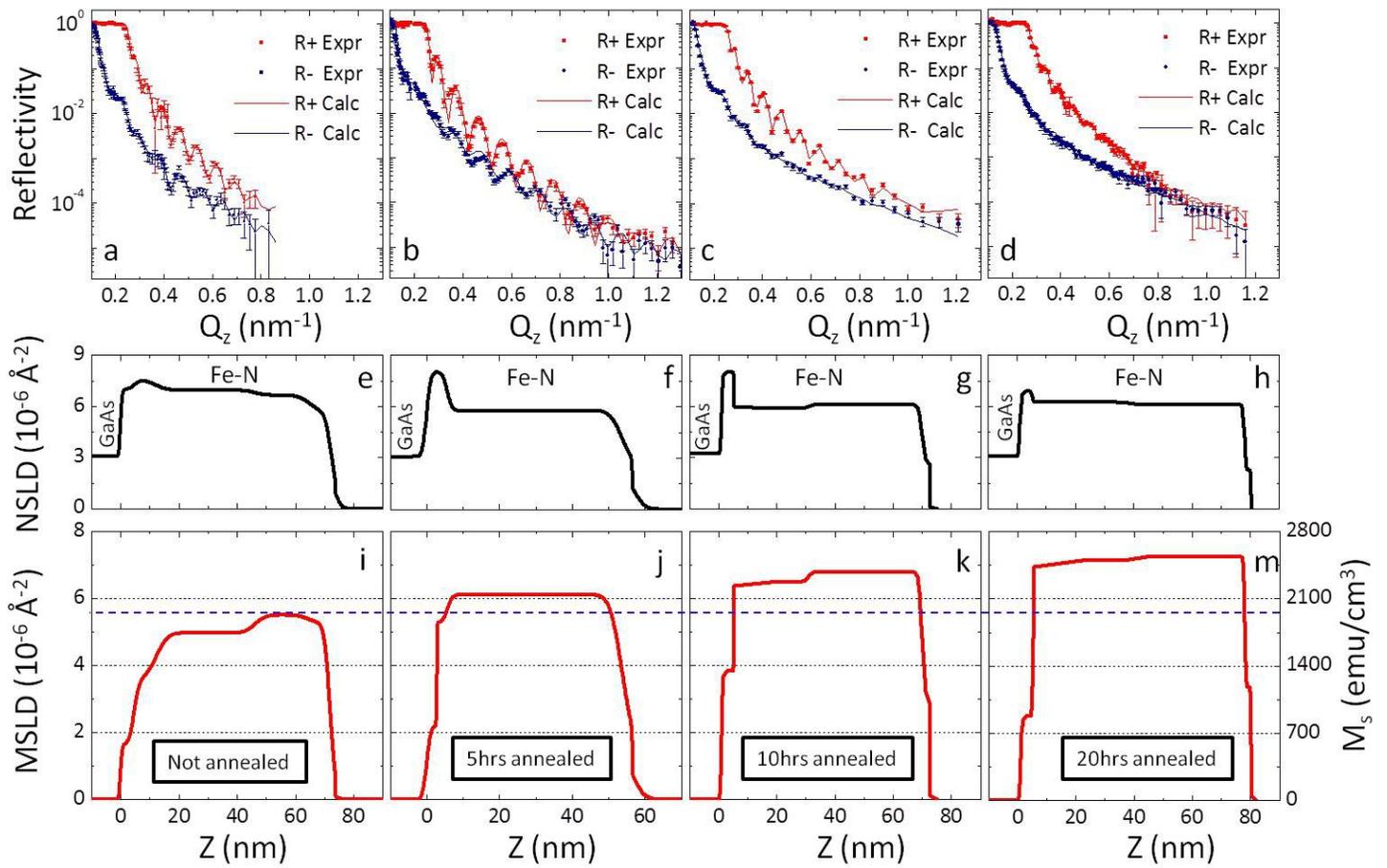

FIG. 5

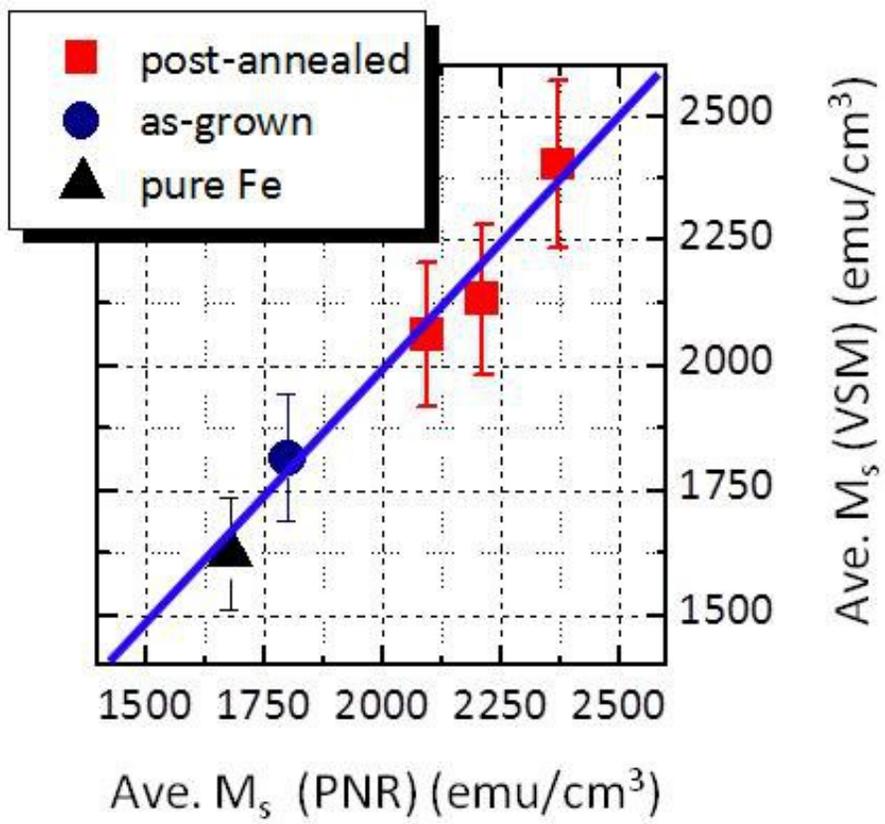

FIG. 6

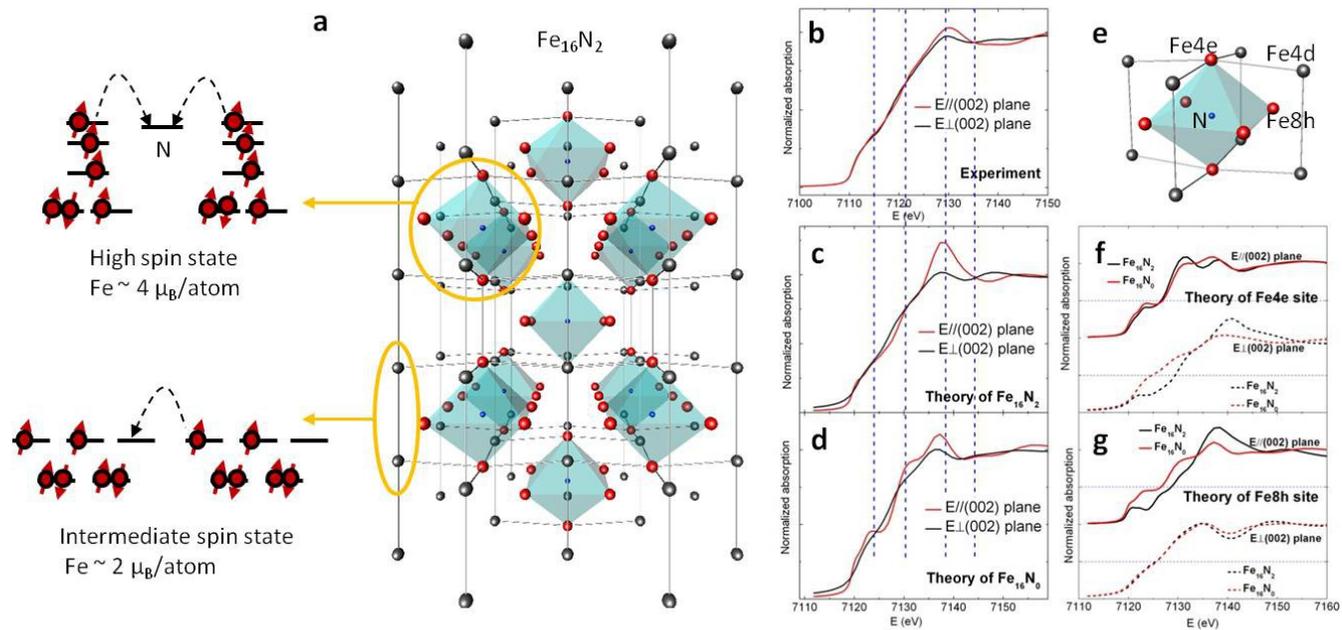

FIG.7

**Reference**

[1] J. C. Slater, J. Appl. Phys. **8**, 385 (1937). L. Pauling, Phys. Rev. **54**, 899 (1938). R. M. Bozorth, Phys. Rev. **79**, 887 (1950).

[2] R. C. O'Handley, Modlern Magnetic Materials Principles and Applications (John Wiley & Sons, Inc., USA, 2000)

[3] We present the saturation magnetization in cgs unit throughout the entire article

[4] T. K. Kim and M. Takahashi, Appl. Phys. Lett. **20** (12), 492 (1972).

[5] Y. Sugita, et. al., J. Appl. Phys. **70** (10), 5977 (1991).

[6] S. Okamoto, et. al., J. Magn. Magn. Mater., 208 102 (2000)

[7] R. M. Metzger, et al., J. Appl. Phys. **76** (10), 6626 (1994)

[8] J. M. D. Coey, J. Appl. Phys. **76** (10), 6632 (1994).

[9] M. Takahashi, et. al., J. Appl. Phys. 79 (8), 5576 (1996).

[10] D. C. Sun, et. al., J. Appl. Phys. **79**, 5440 (1996)

[11] C. Ortiz, G. Dumpich, and A. H. Morrish., Appl. Phys. Lett., **65,** 2737 (1994)

[12] M. A. Brewer, K. M. Krishnan and C. Ortiz, J. Appl. Phys. **79,** 5321 (1996)

[13] N. Ji, et. al, Appl. Phys. Lett. 98, 092506 (2011).

[14] R. Coehoorn, et. al., Phys. Rev. B **48** 3830 (1993).

[15] W Y Lai, Q Q Zheng and W Y Hu, J. Condens. Matter **6** L259 (1994)

[16] G. W. Fernando,et. al., Phys. Rev. B **61** 375 (2000)

[17] A. Sakuma, J. Appl. Phys. **79** 5570 (1996)

[18] S. Asano and M. Yamaguchi, Physica B **237** 541 (1997)

[19] At the annual conference on Magnetism and Magnetic Materials in 1996, a symposium was held on the topic $Fe_{16}N_2$, on which both theorists and experimentalists presented their work. The papers were published in *J. Appl. Phys.* **79** 5564-5581 (1996). No decisive conclusion was drawn on whether it has giant saturation magnetization. Since 1996.

[20] N. Ji, X. Q. Liu, J.-P. Wang, New J. Phys. 12, 063032 (2010)

[21] J. M. D. Coey, Phys. World 6, 25 (1993)

[22] G.F. Felcher, Physica B 192, 137 (1993)

[23] J.A.C. Bland, et al., Phys. Rev. Lett. 58, 1244 (1987)

[24] M. Naoe, et. al., IEEE Magn. Trans. 16, 646 (1980); J. R. Shi, et al, Thin Solid Films, 420, 172 (2002)

[25] N. Ji et al., J. Appl. Phys. 109, 07B767 (2011)

[26] S. L. Qiu, et. al., Phys. Rev. B 64, 104431 (2001)

[27] J. Buschbeck, et al., Phys. Rev. Lett. 103 216101 (2009)

[28] Please see the methods for the sample preparation conditions

[29] H. Rauch and D. Petraschek, in Neutron Diffraction, edited by H.Dachs, Vol. 6 of Topics in Current Physics (Springer, New York, 1978).

[30] M. Born and E. Wolf. Principle of Optics, 6th ed. (Pergamon, Elmsford, NY, 1980).

[31] G.P. Felcher, R.O. Hilleke, R.K. Crawford, J. Haumann, R. Kleb, G. Ostrowski, Rev. Sci. Instrum. 58, 609 (1987); J.A.C. Bland, B. Heinrich (Eds.) Ultrathin Magnetic Structures, Springer, Berlin,Heidelberg, 1994

[32] The MSLD ($\rho b_m$) is the same physical parameter as the saturation magnetization ($M_s$), both of which are a measure of the magnetic density. In cgs unit, the conversion formulism is $3.5 \times 10^8$ x $\rho b_m$ ($\text{Å}^{-2}$) = $M_s$ (emu/cm$^3$)

[33] V. Lauter, H. Ambaye, R. Goyette, W.-T. H. Lee, A. Parizzi, Physica B 404, 2543 (2009); V. Lauter-Pasyuk, J. Phys. IV France 1 (2007); V. Lauter-Pasyuk, H. J. Lauter, B. P. Toperverg, L. Romashev, and V. Ustinov, Phys. Rev. Lett. 89, 167203 (2002)

[34] D. E. Sayers et al., Phys. Rev. Lett. 27, 1204 (1971)

[35] J.J. Rehr et al., Rev. Mod. Phys 72, 621 (2000)

[36] J. Stöhr, NEXAFS Spectroscopy, Springer (1992)

[37] A. Bianconi et al., Phys. Rev. B 26, 6502 (1982)

[38] A. L. Ankudinov, et al., Phys. Rev. B 67, 115120 (2003).

[39] To construct the $Fe_{16}N_2$ crystal, a space group of I4/mmm and experimental lattice constants were used. a cluster of atoms with a size of 5.47 Å is applied for the calculation of full-multiple scattering (FMS), a cluster of atoms with a size of 4.0 Å is used for the calculation of the self-consistent-field (SCF) muffin-tin atomic potential, the Hedin–Lundqvist self-energy is used by default. To calculate the XANES of $Fe_{16}N_0$, the crystal structure and lattice constant were chosen to be identical to single crystal $Fe_{16}N_2$. The sites that originally occupied by N in $Fe_{16}N_2$ were left empty on purpose when constructing the dummy structure $Fe_{16}N_0$.